\documentclass[aip, preprint]{revtex4-1}
\usepackage{graphicx}
\usepackage{siunitx}
\usepackage{ulem}
\usepackage{textcomp} 
\usepackage{amsmath}

\begin{document}

\title{The Onset of Phase Separation in the Double Perovskite Oxide La$_2$MnNiO$_6$}
	
\author{Steven R. Spurgeon}
\affiliation{Physical and Computational Sciences Directorate, Pacific Northwest National Laboratory, Richland, WA 99352, USA}

\author{Peter V. Sushko}
\affiliation{Physical and Computational Sciences Directorate, Pacific Northwest National Laboratory, Richland, WA 99352, USA}

\author{Arun Devaraj}
\affiliation{Physical and Computational Sciences Directorate, Pacific Northwest National Laboratory, Richland, WA 99352, USA}

\author{Yingge Du}
\affiliation{Physical and Computational Sciences Directorate, Pacific Northwest National Laboratory, Richland, WA 99352, USA}

\author{Timothy Droubay}
\affiliation{Physical and Computational Sciences Directorate, Pacific Northwest National Laboratory, Richland, WA 99352, USA}

\author{Scott A. Chambers}
\affiliation{Physical and Computational Sciences Directorate, Pacific Northwest National Laboratory, Richland, WA 99352, USA}
\email{sa.chambers@pnnl.gov}

\begin{abstract}

Identification of kinetic and thermodynamic factors that control crystal nucleation and growth represents a central challenge in materials synthesis. Here we report that apparently defect-free growth of La$_2$MnNiO$_6$ (LMNO) thin films supported on SrTiO$_3$ (STO) proceeds up to 1--5 nm, after which it is disrupted by precipitation of NiO phases. Local geometric phase analysis and ensemble-averaged X-ray reciprocal space mapping show no change in the film strain away from the interface, indicating that mechanisms other than strain relaxation induce the formation of the NiO phases. \textit{Ab initio} simulations suggest that oxygen vacancies become more likely with increasing thickness, due to the electrostatic potential build-up associated with the polarity mismatch at the film-substrate interface; this, in turn, promotes the formation of Ni-rich regions. These results suggest that the precipitate-free region could be extended further by increasing the oxygen chemical potential through the use of an elevated oxygen pressure or by incorporating electron redistributing dopants to suppress the built-in potential.

\end{abstract}

\maketitle


The formation of undesirable and uncontrolled phases during thin film nucleation and growth represent a fundamental obstacle to the atomically-precise synthesis of materials with targeted properties. As highlighted in recent reviews,\cite{DeYoreo2016, DeYoreo2015} it is extremely difficult to observe and harness kinetic processes, hindering our efforts to control film growth.\cite{Sear2016} The conditions associated with different techniques, such as pulsed laser deposition (PLD) and molecular beam epitaxy (MBE), lead to vastly different energy landscapes that govern the synthesis process. Furthermore, it is understood that film deposition is a dynamical process---namely, conditions at the growth front change during deposition resulting in complex synthesis outcomes. There is a pressing need to understand the factors that govern this dynamical behavior and how they can trigger the formation of inhomogeneities, from isolated defects to parasitic phases.\cite{Kaiser2002, Dietl2010, Kalinin2010}

Coherently strained epitaxial thin films offer a highly controlled environment in which to identify the signatures and guiding mechanisms of phase separation events. The ability of perovskite oxides to phase separate has already been exploited to produce complex nanostructured materials;\cite{Zhang2014, He2011, Wang2010, Macmanus-Driscoll2010, Macmanus-Driscoll2008} however, predictive control of nanocomposites remains elusive, and more insight is needed into the details of elementary processes and species that mediate phase separation. For example, the polarity mismatch at polar / non-polar interfaces is thought to induce NiO phase separation in LaNiO$_3$ / LaAlO$_3$ superlattices grown on SrTiO$_3$ (STO).\cite{Detemple2011} Lazarov \textit{et al.} have demonstrated that polar Fe$_3$O$_4$ / MgO (001) interfaces can be stabilized through the formation of Fe nanocrystals, rather than through interface reconstruction, faceting, or intermixing.\cite{Lazarov2003} These studies illustrate that oxide thin film systems form through complex synthesis routes dictated by chemical composition, as well as external conditions.

Recently, we have characterized Ni-rich precipitates in the double perovskite La$_2$MnNiO$_6$ (LMNO), which our \textit{ab initio} simulations indicate form in an oxygen-poor growth environment.\cite{Spurgeon2016} The formation of defects in oxygen-deficient conditions has also been noted by Guo \textit{et al.};\cite{Guo2008} interestingly, the authors found that PLD-deposited LMNO films in 100 mTorr pO$_2$ exhibited a defect-free 5--10 nm interface layer atop which a defective film formed. The authors proposed that homogeneous film growth proceeds until a critical thickness, beyond which defect formation acts to relax the relatively small 1\% lattice mismatch between the film and substrate; however, no evidence of misfit dislocations was found and no conclusive phase separation mechanism was identified. In contrast to the simple perovskite structure, the presence of two different $B$-site cations in the double perovskite can lead to widespread anti-site defects,\cite{Saha-Dasgupta2012, Singh2009} which can in turn favor oxygen vacancy formation. The interaction of cation species and oxygen vacancies provides a potentially unique way for double perovskites to screen interface charge and represents an unexplored driver for phase separation.

Here we report that, while synthesis conditions favor the formation of NiO precipitates inside LMNO films,\cite{Spurgeon2016} there is a 1--5 nm region at the LMNO / STO interface that is free of these defects. We then propose an atomic-scale mechanism that triggers the formation of Ni-rich regions outside of this near-interface region and suggest how precipitate morphology may be controlled by dynamically changing synthesis conditions.

\begin{figure}
\includegraphics[width=\textwidth]{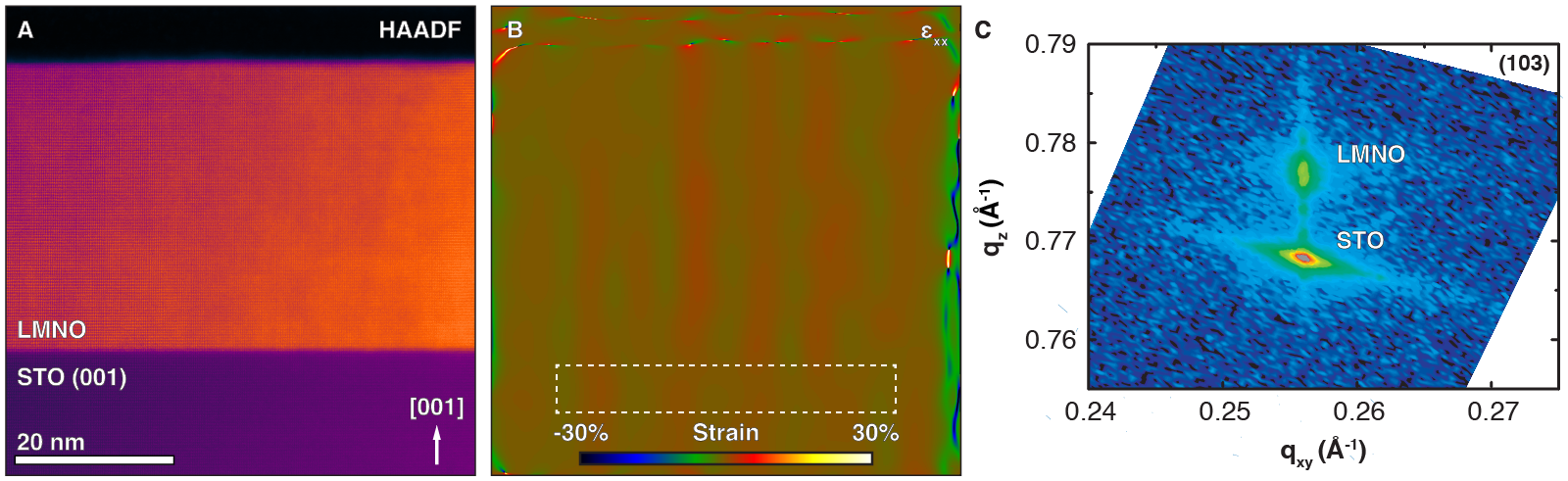}
\caption{Measurement of film quality and strain state. (A) Colorized cross-sectional STEM-HAADF image and corresponding (B) in-plane strain ($\varepsilon_\textrm{xx}$) map generated using GPA. The dashed box marks the reference region. (C) X-ray reciprocal space map around the STO (103) reflection showing that the LMNO layer is coherently strained in-plane.\label{stem_overview}}
\end{figure}

We have prepared several 40 nm-thick La$_2$MnNiO$_6$ films on STO (001) substrates using molecular beam epitaxy (MBE), as described elsewhere.\cite{Spurgeon2016} Figure \ref{stem_overview}.A shows a cross-sectional high-angle annular dark field (STEM-HAADF) image of the sample, confirming a heteroepitaxial film structure with a sharp film-substrate interface. While the overall contrast in this region is uniform, we observe some contrast variations that indicate possible composition fluctuations. We have previously determined nanoscale NiO precipitate formation to be the origin of the contrast variations, but the onset of phase separation is unclear.\cite{Spurgeon2016} To assess the local strain state of the film, we perform geometric phase analysis (GPA), which allows us to measure local strain at nanometer-scale spatial resolution with $\sim0.1\%$ strain resolution directly from STEM images.\cite{Hytch2014, Hytch2007} Figure \ref{stem_overview}.B shows a map of the in-plane strain ($\varepsilon_\textrm{xx}$) component, which exhibits a near constant value with only minor fluctuations. These local strain measurements show no indication of overall film strain relaxation, a finding present in multiple GPA maps and further supported by ensemble-averaged X-ray reciprocal space mapping (RSM). As shown in Figure \ref{stem_overview}.C, the film is coherently strained to the substrate, with no relaxation within the resolution of our measurements. These results suggest that the formation of NiO precipitates is driven by factors other than strain relaxation and that further analysis of secondary phase morphology is needed to understand the nucleation process.

\begin{figure}
\includegraphics[width=\textwidth]{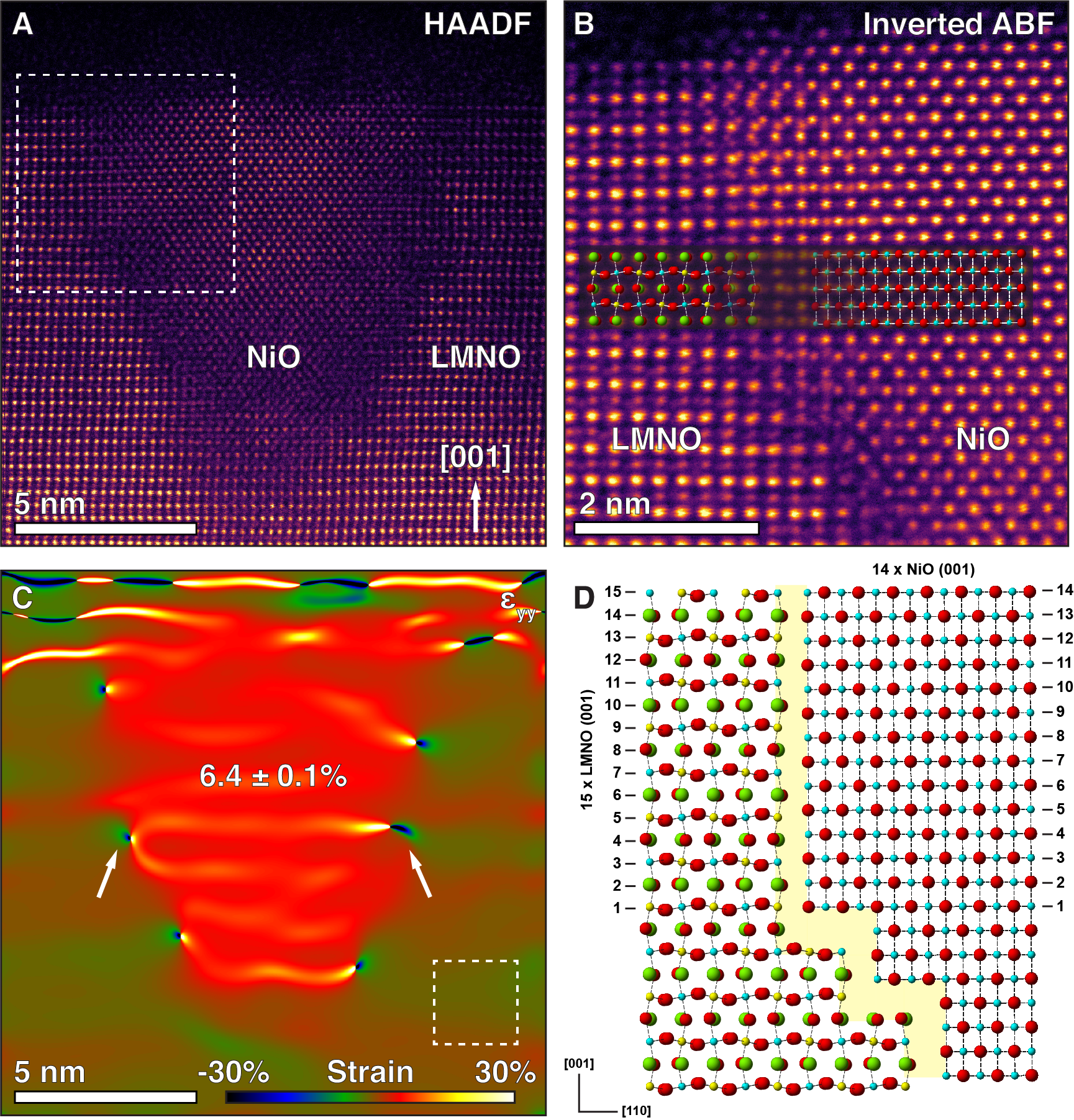}
\caption{Interface between the secondary phase and matrix. (A) Cross-sectional STEM-HAADF image of a thin sample region taken along the [110] zone-axis, in which an isolated precipitate is clearly visible. (B) Non-rigid registered inverted ABF image of the dashed in region in (A), overlaid with theoretical LMNO and NiO crystal models. (C) Out-of-plane strain ($\varepsilon_\textrm{yy}$) map generated using GPA, with misfit dislocations at the LMNO / NiO interface indicated by arrows. The dashed box indicates the reference region. (D) Model of lattice matching across the interface, showing how the misfit between the two phases is accommodated by dislocations. Atoms: Red = O, green = La, cyan = Ni, and yellow = Mn. \label{stem_particle}}
\end{figure}

Figure \ref{stem_particle}.A shows a high-angle annular dark field (STEM-HAADF) image from a very thin region of the foil in which a single NiO precipitate is present. This image reveals the complex spatial distribution and morphology at the surface of the nanocomposite structure. First note the inverted ``pyramid''-like shape of the NiO precipitate, which progresses from a 2--3 nm base to a $\sim$10~nm wide mouth at the film surface. While columnar structures have recently been observed in other perovskites,\cite{Zhang2016} this unique morphology is shaped by facets along $<$111$>$-type directions, inclined 50--53$^{\circ}$ to the substrate, compared to the theoretical angle of 54.7$^{\circ}$ between the (111) and (001) planes. We find that both the LMNO matrix and the NiO secondary phase are crystalline: the former possesses a $P2_1/n$ double perovskite structure, while the latter has a $Fm\bar{3}m$ rock-salt structure. Moreover, there appears to be a distinct quasi-epitaxial $[110]_\textrm{NiO}$ // $[110]_\textrm{LMNO}$ $\vert \vert$ $(001)_\textrm{NiO}$ // $(001)_\textrm{LMNO}$ relationship between the phases. More detail of the interface between the precipitate and matrix is shown in Figure \ref{stem_particle}.B, which presents an inverted annular bright field (STEM-ABF) image that is sensitive to light elements in the structure. This unfiltered image is the result of non-rigid alignment, which improves signal-to-noise and reduces scan artifacts that can obscure the underlying data (see methods for details). Theoretical crystal structures have been overlaid onto this structure, showing the excellent lattice match between the phases, which are separated by a transition region likely resulting from tapering of the crystal in the beam direction.

Let us consider the epitaxial out-of-plane strain between the matrix and secondary phase. Assuming bulk lattice parameters of $c_\textrm{LMNO} = 0.387$ nm and $c_\textrm{NiO} = 0.417$ nm,\cite{Hashisaka2006, Fievet1979} we expect a large out-of-plane lattice mismatch of,
\begin{equation*}
\frac{c_\textrm{NiO} - c_\textrm{LMNO}}{c_\textrm{NiO}} = \frac{0.417 - 0.387}{0.417} = 7.2\%
\end{equation*}

\noindent Based on the behavior of other perovskite-based nanocomposites,\cite{Macmanus-Driscoll2008} we expect that misfit dislocations will form to accommodate this strain. Figure \ref{stem_particle}.C shows an array of misfit edge dislocations decorating the NiO / matrix interface. This figure shows a colormap of local out-of-plane strain ($\varepsilon_\textrm{yy}$) relative to the bulk matrix strain state; in the center of the NiO precipitate we observe a clear out-of-plane expansion of $6.4 \pm 0.1\%$, suggesting that the majority of the lattice mismatch is relaxed by dislocations. From our GPA maps we estimate an average dislocation spacing of $\sim$3.2~nm, which we can use to calculate the lattice match between the phases. We note that the closest (001) plane spacing for each phase is $d_\textrm{NiO (001)} = 0.2085$ nm and $d_\textrm{LMNO (001)} = 0.1934$ nm. Assuming no deviation from this bulk spacing, the best matching combination is 14 (001) planes in NiO and 15 (001) planes in LMNO. This combination yields total distances of 2.92 and 2.90~nm in NiO and LMNO, respectively, in good agreement with the measured dislocation spacing of $\sim$3.2~nm. Despite the difference in out-of-plane lattice parameters, the LMNO films are fully coherent with the STO substrates even with the inclusion of NiO phase, as revealed by reciprocal space maps taken along the (103) film reflection.\cite{Spurgeon2016} Using this information, we are able to construct a model of the interface lattice matching, as shown in Figure \ref{stem_particle}.D. The clear boundary between two crystalline regions shows that large-scale phase ordering of tens of nanometers is possible in this system and that a distinct orientation relationship can be preserved between the constituent phases far from the substrate.

\begin{figure}[t]
\includegraphics[width=\textwidth]{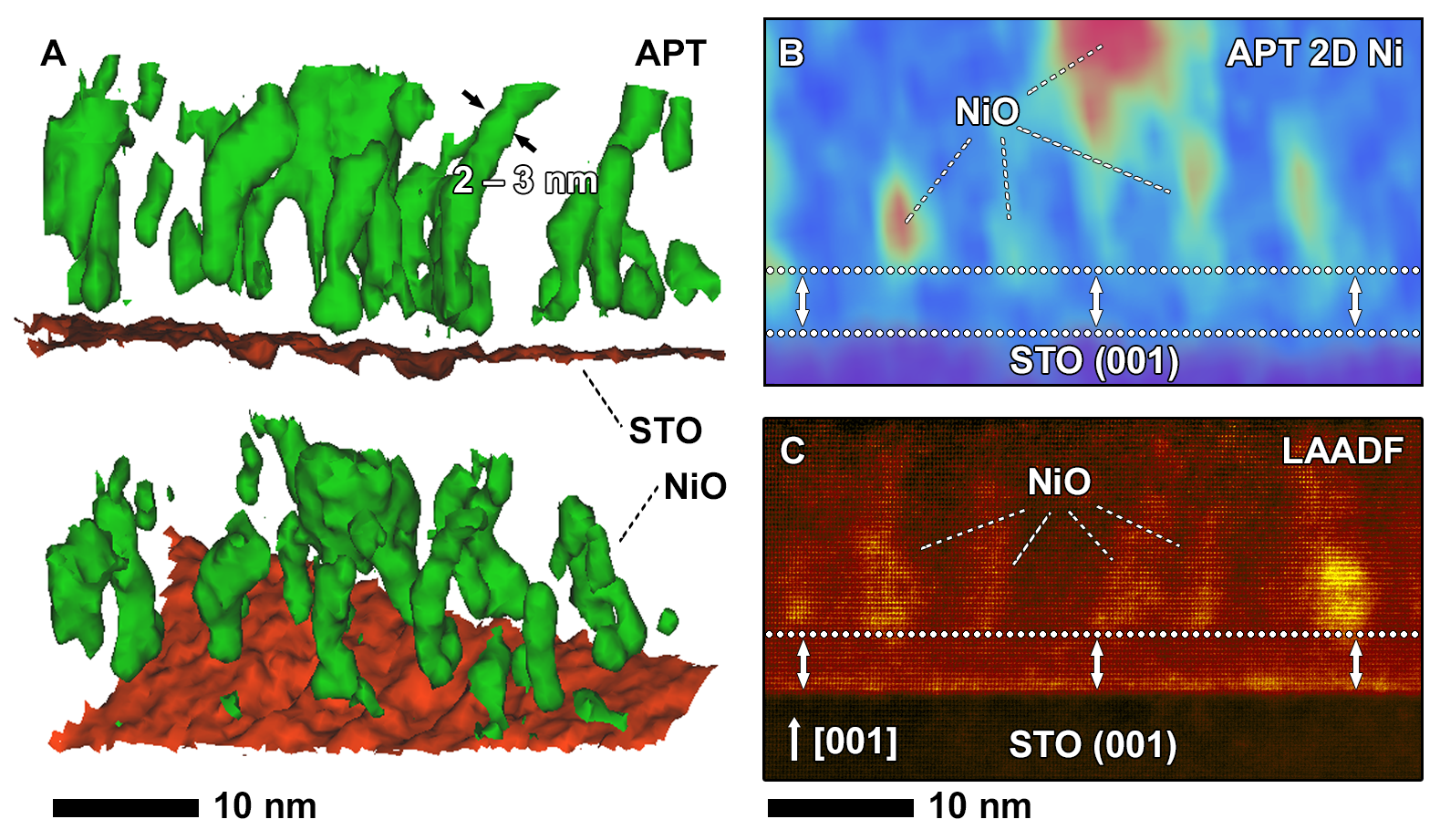}
\caption{Multidimensional APT / STEM analysis of secondary phase morphology. (A) Transverse (top) and oblique (bottom) views of a 15 at \% Ni iso-composition surface of the sample, revealing faceted NiO precipitates marked by arrows. (B) 2D Ni composition map from the dataset in (A), highlighting the gap between precipitates and the substrate (marked by arrows). Volume: 1 nm $\times$ 30 nm $\times$ 40 nm. (C) STEM-LAADF image taken along the [110] zone-axis, emphasizing strain around the NiO precipitates.\label{apt}}
\end{figure}

With an overall structural understanding of the NiO / matrix interface, we now discuss the mechanism of NiO precipitate formation. While STEM imaging provides valuable insight into local morphology, projection issues make it difficult to extract three-dimensional (3D) compositional information that can inform our models. We therefore performed atom probe tomography (APT) measurements, which allow us to reconstruct the 3D spatial distribution of individual elements at the nanometer scale.\cite{Devaraj2017} Figure \ref{apt}.A shows a volume reconstruction of LMNO / STO interface, with a 15 at \% Ni iso-composition surface shown in green, highlighting NiO columns, and the film / STO interface shown in red. While this construction allows us to sharply delineate the Ni-rich regions of the film, we note that there is a composition gradient from the core of each precipitate to the matrix. Electron energy loss spectroscopy (STEM-EELS) measurements, presented in Figure S2, confirm this gradient and also reveal a Mn enrichment around the NiO phases. We observe a strikingly homogeneous distribution of columnar NiO precipitates approximately 2--3 nm in diameter, running from 1--5 nm above the substrate to the surface. Many of these structures exhibit pinched, neck-like features resulting from faceting, as indicated by the arrows in Figure \ref{apt}.A, as well as inverted pyramid-like shapes similar to those in Figure \ref{stem_particle}.A. This behavior is comparable to the branched growth previously observed in NiFe$_2$O$_4$ / LaFeO$_3$ driven by energy-reducing facet formation.\cite{Comes2017}

Most interestingly, we observe a region in the immediate vicinity of the substrate that is largely devoid of significant phase separation (\textit{i.e.} ordered LMNO). This behavior contrasts markedly with related systems, such as La$_{0.67}$Ca$_{0.33}$MnO$_3$ / MgO\cite{Lebedev2002} and LaNiO$_3$ / LaAlO$_3$,\cite{Detemple2011} in which secondary phases nucleate directly at the film / substrate interface. A clearer representation of this interface region is shown in the two-dimensional (2D) Ni composition map in Figure \ref{apt}.B, constructed by taking a 2D volume of 1 nm $\times$ 30 nm $\times$ 40 nm through a representative part of the APT dataset. This figure indicates that the upper portion of the film is dominated by high aspect-ratio NiO precipitates that are missing from a 1--5 nm layer at the substrate. Low-angle annular dark field (STEM-LAADF) measurements, shown in Figure \ref{apt}.C, confirm this observation. This imaging mode is highly sensitive to a strained layer at the film-substrate interface, as well as the presence of large strains around the columnar NiO precipitates in the upper portion of the image; the core of the precipitates are largely absent from a layer immediately adjacent to the substrate. Taken together, these results suggest that the initial film growth proceeds in a homogeneous fashion and that the onset of phase separation does not occur until 1--5 nm film thickness is reached.

These observations point to a growth mechanism that depends on the LMNO film thickness and which can be affected by, for example, build-up of lattice strain and/or build-up of the electrostatic potential due to the polar mismatch at the interface. While strain relaxation \textit{via} misfit dislocations and other defects is commonly encountered in oxide thin films, our GPA and XRD results suggest that such relaxation is minor. Anti-site defect clustering has been explored in the LiMnPO$_4$ and LiFePO$_4$ systems; while isolated defects were energetically preferred in the former, the latter exhibited a significant free energy reduction associated with the formation of defect clusters. Similar defect clustering could induce phase separation in the present study. Previous study of LaNiO$_3$ / LaAlO$_3$ interfaces has also raised the intriguing possibility of the phase separation driven by a polar mismatch at the film-substrate interface.\cite{Detemple2011}

To reveal the atomic-scale origin of LMNO's properties in the initial growth stages, we have evaluated the thermodynamic stability of several types of defect structures as a function of LMNO film thickness using the density functional theory (DFT) formalism. We note that idealized LMNO / STO interface is a polar / nonpolar junction that is expected to exhibit electrostatic potential build-up in the LMNO film as its thickness ($n$) increases. In contrast to the well-known case of LaAlO$_3$ / SrTiO$_3$ (LAO / STO), Ni and Mn species in LMNO have partially occupied 3$d$ shells, which can facilitate electron charge redistribution within the film and, thereby offsetting electrostatic potential build up without cross-interface cation intermixing.\cite{Qiao2011, Chambers2010}

\begin{figure}
\includegraphics[width=\textwidth]{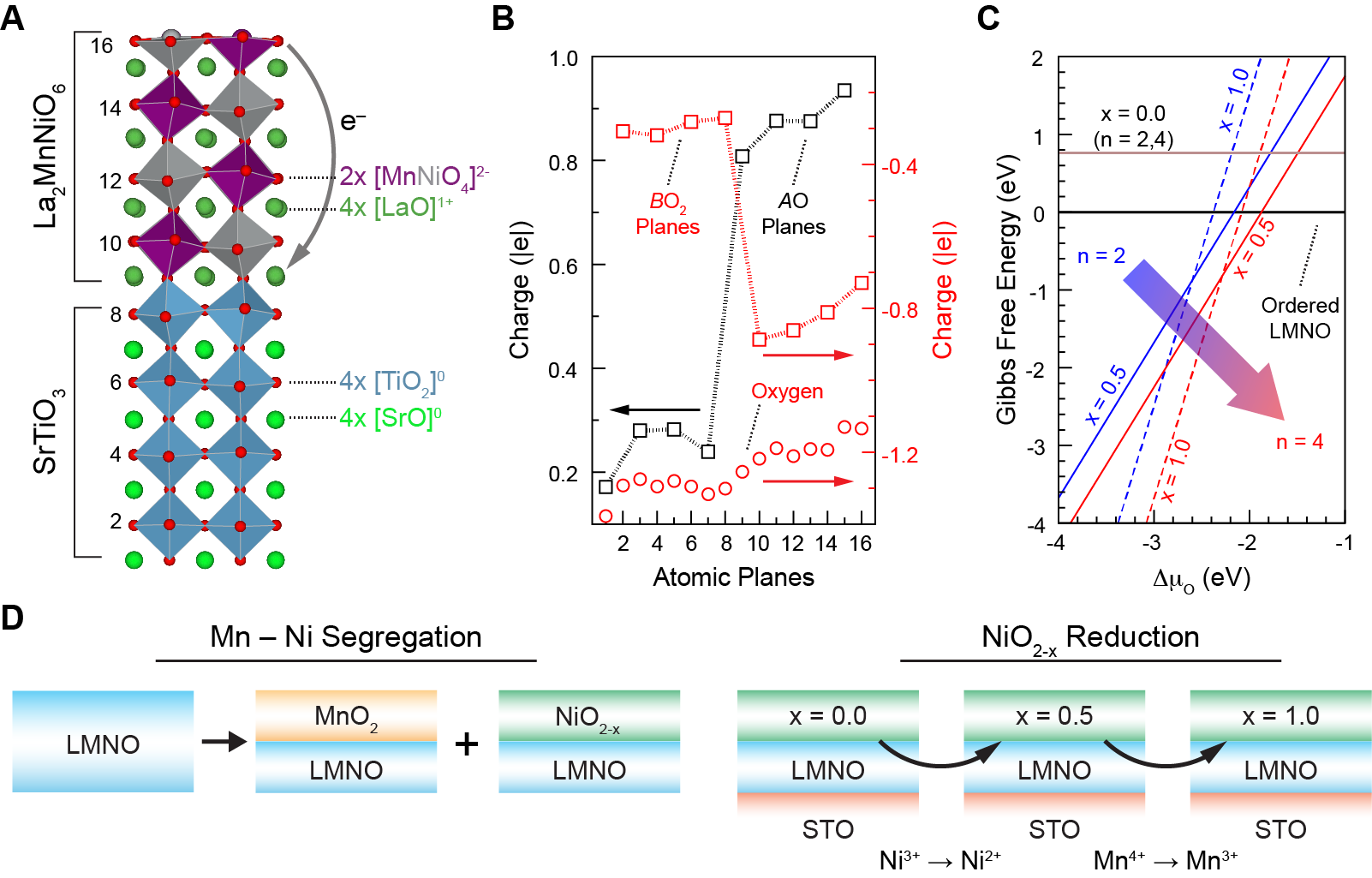}
\caption{Charge redistribution in LMNO facilitates the formation of NiO precipitates. 
(A) Periodic slab used in \textit{ab initio} modeling of defect formation mechanisms. Thickness of the LMNO  film was varied between $n=1$ and $n=4$. Numbers on the left indicate atomic planes. 
(B) Charges of the $A$O and $B$O$_2$ planes (squares) per 1$\times$1 lateral cell and average oxygen charge (circles) in ordered stoichiometric LMNO ($n=4$) show that electrons transfer from the surface to LMNO / STO interface. 
(C) Gibbs free energies for LMNO ($n=2,4$), in which the surface $B$O$_2$ plane is segregated into MnO$_2$ and NiO$_{2-x}$ regions for $x=0.0$, 0.5, and 1.0, show that phase separation of NiO becomes preferred as oxygen chemical potential decreases and $n$ increases.
(D) Proposed mechanism of initial stages of NiO segregation: formation of Ni-rich areas is facilitated by $V_{\text O}$ and by the ability of LMNO to accommodate additional charge in the form of Mn$^{3+}$ ions (see also Figures S4--6).\label{DFT}}
\end{figure}

According to our calculations, the electrostatic field in the film is $\sim$0.1~V $\textrm{\AA}^{-1}$ (see Figure S4.A for more details) for $n=4$~u.c., which is significantly smaller than the corresponding field of $\sim$0.24~V $\textrm{\AA}^{-1}$ theoretically predicted for idealized LAO / STO.\cite{Lee2008} The low field value is consistent with the following charge redistribution scenario: as the LMNO thickness increases from $n=2$ to $n=4$~u.c., $\sim$0.3 $|e|$ transfers from the top-most Mn$_{2}$Ni$_{2}$O$_8$ plane to the positively charged interface (Mn$_{0.5}$Ni$_{0.5}$O$_2$ / LaO / TiO$_2$) and partially offsets the internal field induced by the polarity mismatch, as shown in Figures \ref{DFT}.A,B. In turn, as the top-most LMNO plane becomes positively charged with respect to the bulk lattice, the surface anions become destabilized. This effect is illustrated in Figure S5, which shows that, with the exception of a single monolayer LMNO film, the energy cost of forming an oxygen vacancy in the top-most LMNO plane decreases with increasing film thickness. We find that two electrons associated with such oxygen vacancies are primarily localized at the Mn species located near the vacancy site; the resulting electron density redistribution and atomic displacements further suppress the internal field to less than 0.05~V $\textrm{\AA}^{-1}$ (see Figure S4.B).

Next we demonstrate that cation (Mn and Ni) site disorder is strongly correlated to the local oxygen deficiency. First we note that Mn / Ni substitutional defects have relatively low formation energies. For example, swapping neighboring Mn and Ni atoms in the top-most LMNO plane results in the formation of Mn and Ni rows oriented along [100] lattice vectors. The cost of forming this configuration is $\sim$0.3~eV with respect to the ordered LMNO film. Similarly, the structure formed by off-register deposition of two consecutive LMNO unit cells is only 0.3~eV less stable than the corresponding ordered structure (see Figure S6 for more details). Assuming that the formal ionic charges of Ni, Mn, and O species remain unchanged, such non-ordered configurations have local regions that are either negatively (Ni$^{2+}$O$^{2-}$Ni$^{2+}$) or positively (Mn$^{4+}$O$^{2-}$Mn$^{4+}$) charged with respect to equivalent regions in the ideal lattice (Ni$^{2+}$O$^{2-}$Mn$^{4+}$). Oxygen vacancies can stabilize such Ni-rich and Mn-rich regions simultaneously. Indeed, if the vacancy is located between two Ni sites and the two electrons associated with this vacancy localize on the Mn species, then the local charges of both Ni-rich (Ni$^{2+}V_{\text O}$Ni$^{2+}$) and Mn-rich (Mn$^{3+}$O$^{2-}$Mn$^{3+}$) regions become equivalent to that in the ordered lattice. We note that similar association of oxygen vacancies and Ni-rich regions was proposed to take place in LiNi$_{0.5}$Mn$_{1.5}$O$_{4-\delta}$ spinel and control voltage suppression in Li-ion batteries that use this material.\cite{Sushko2013}

To quantify the link between oxygen deficiency and Mn / Ni disorder, we simulate LMNO films in which the surface plane is segregated into pure MnO$_2$ and NiO$_2$; we then calculate the stability of this segregated system as a function of oxygen content for several different film thicknesses. Given the size of the lateral cell used in this work, the composition of the fully oxidized top plane in the Ni-rich case is Ni$_4$O$_8$, which corresponds to the average formal charge of 3+ for the Ni species. Hence, forming two oxygen vacancies in this plane, which corresponds to Ni$_4$O$_6$ composition, converts all Ni species to Ni$^{2+}$ ions. Forming two more vacancies results in a composition equivalent to NiO (planar Ni$_4$O$_4$) and produces four electrons that can localize either between Ni$^{2+}$ ions in the NiO$_{2-x}$ plane (similar to the electrons in the $F$-center in MgO) or on Mn and Ti 3$d$-states. The full set of the calculated vacancy formation energies in the Ni-rich LMNO is given in Table SI. Importantly, formation of the first two $V_{\text O}$, \textit{i.e.} Ni$^{3+} \rightarrow$Ni$^{2+}$ conversion, requires a relatively low energy (1.1--2.2 eV) that is almost independent of the LMNO film thickness. In contrast, formation of the second two $V_{\text O}$, \textit{i.e.} (Mn,Ti)$^{4+}$ $\rightarrow$ (Mn,Ti)$^{3+}$ conversion, requires a higher energy (2.0--3.2 eV) that decreases slowly with increasing film thickness.

Figure \ref{DFT}.C shows the Gibbs free energies calculated as a function of oxygen chemical potential $\Delta \mu_{\text O}$\cite{Heifets2007} for the cases of $n=2$ and $n=4$ u.c. and $x=0.0$, 0.5, and 1.0, \textit{i.e.}, up to four $V_{\text O}$ per 2$\times$2 lateral cell. The experimental LNMO / STO deposition conditions ($T=650$$^{\circ}$C and $p_{{\text O}2} \sim 1 \times 10^{-5}$ Torr) correspond to a $\Delta \mu_{\text O}$ value of $\sim$ --1.3 eV. It is clear that the combination of segregated MnO$_2$ and oxygen deficient NiO$_{2-x}$ (up to $x \approx$ 0.5) regions at the LMNO surface becomes more stable as the LMNO thickness increases. We attribute this effect to the increasing number of Mn$^{4+}$ ions, which can be converted to Mn$^{3+}$ (in general Mn$^{3+\gamma}$, $\gamma>$0) ions and offset the cost associated with vacancy formation, as illustrated in Figure \ref{DFT}.D.

Our experimental measurements reveal the presence of a phase-pure 1--5 nm interface region of LMNO that is devoid of NiO precipitates. XRD and GPA confirm that the film is uniformly strained, pointing toward another mechanism for phase separation. These findings help clarify anomalous features in prior TEM and electron diffraction studies,\cite{Guo2008, Singh2007} which we believe are the result of nanodomain formation that is not apparent in conventional XRD measurements. We propose that when the LMNO is thin, the polarity discontinuity potential is low and its effect on the formation of oxygen vacancies and cation disorder is negligible. While we cannot rule out the formation of these defects, the relatively high oxygen vacancy formation energies for thin films (see Table SI) and the experimentally observed phase-pure LMNO within 1--5 nm near the interface, suggest that the concentration of these vacancies is low. As the film grows, the built-in potential arising from the polar discontinuity at the LMNO / STO interface becomes sufficient to promote the formation of surface oxygen vacancies and associated Mn / Ni site disorder, including segregation of Mn-rich and Ni-rich areas. Furthermore, as the number of Mn$^{4+}$ ions in the film becomes sufficiently large, additional vacancies can form as discussed above and promote the formation of Ni-rich regions with a chemical composition equivalent to that of NiO.


Our measurements provide direct evidence that the onset of NiO phase separation in LMNO / STO films begins 1--5 nm from the interface. We note that local GPA and ensemble-averaged X-ray RSM indicate that the LMNO film is fully strained to the substrate and that no measurable relaxation takes place. \textit{Ab initio} modeling suggests that Ni-rich areas form in response to oxygen-deficiency; in turn, the likelihood of oxygen vacancy formation increases with increasing LMNO thickness due to the polarity mismatch at the LMNO / STO interface. Beyond a 1--5 nm phase-pure interface region, NiO precipitates form in a quasi-epitaxial growth relationship to the substrate, indicating that synthesis in an environment of limited oxygen fugacity may serve as a route to design ordered nanocomposites. As defect-free LMNO is deposited atop LMNO containing NiO precipitates, the polarity mismatch between these regions and the resulting built-in electrostatic potential will further enhance oxygen vacancy formation; this, in turn, will trigger the precipitation of new Ni-rich regions.

Finally, we propose that dynamic tuning of growth conditions can be used to control the onset of NiO phase separation. In particular, increasing oxygen chemical potential either by elevating the growth pressure or by introducing additional oxygen atmosphere annealing steps can suppress the formation of oxygen vacancies, even in the presence of built-in electrostatic potential. Alternatively, doping LMNO with transition metals that enable efficient electron redistribution within the film would also suppress the built-in potential, as well as its effect on the oxygen vacancy formation. In both cases it may be possible to dictate the onset, distribution, and morphology of the NiO precipitates, paving a way for control of nanoscale phase separation and precise atomic-scale synthesis.

\clearpage

\section*{Supplementary Material}

Supplementary material containing details on the methods used, STEM imaging, EELS, APT, and DFT is available online.

\section*{Acknowledgements}

S.R.S. thanks Drs. Yuanyuan Zhu and Quentin Ramasse for constructive discussions. This work was supported by the U.S. Department of Energy, Office of Science, Division of Materials Sciences and Engineering under award \#10122. All work was performed as part of a science theme user proposal in the Environmental Molecular Sciences Laboratory, a national science user facility sponsored by the Department of Energy's Office of Biological and Environmental Research and located at Pacific Northwest National Laboratory. Pacific Northwest National Laboratory (PNNL) is a multiprogram national laboratory operated for DOE by Battelle.

\clearpage

\bibliography{References}

\end{document}